\documentclass[prl,preprintnumbers]{revtex4}
\usepackage{dcolumn}
\usepackage{bm}
\usepackage{ulem}
\usepackage{epsfig}
\usepackage{epstopdf}
\usepackage{amsmath}
\usepackage{graphicx}
\usepackage{mathrsfs}
\usepackage{amssymb}
\usepackage{amsfonts}
\usepackage{hyperref}
\usepackage{color,soul}
\usepackage{float}
\usepackage{enumerate}
\usepackage{subfigure}
\usepackage{xcolor}
\usepackage{amsthm}
\usepackage{braket}
\usepackage{multirow}    % multiculumn multirow
\usepackage{array}     % column centering
\newcommand{\inner}[2]{\mbox{$\left\langle #1 | #2 \right\rangle$}}

\begin{document}

\title{Towards practical high-speed high dimensional quantum key distribution using partial mutual unbiased basis of photon's orbital angular momentum }
\author{Fumin Wang$^{1^\dag}$}
\author{Pei Zeng $^{2^\dag}$}
%\author{Xiongfeng Ma$^{2^\ddag}$}
\author{Xiaoli Wang$^{1}$}
\author{Hong Gao$^{1}$}
\author{Fuli Li$^{1}$}
\author{Pei Zhang$^{1}$}
\email{zhangpei@mail.ustc.edu.cn}

\affiliation{$^{1}$ Key Laboratory of Quantum Information and Quantum Optoelectronic Devices, Shaanxi Province, Xi'an Jiaotong University, Xi'an 710049, China\\
$^{2}$ Center for Quantum Information,  Institute for Interdisciplinary Information Sciences, Tsinghua University, Beijing 100084, China\\
$^\dag$These authors contributed equally to this work.\\
E-mail: $^*$ zhangpei@mail.ustc.edu.cn}

\begin{abstract}
Quantum Key Distribution (QKD) guarantees the security of communication with quantum physics. Most of widely adopted QKD protocols currently encode the key information with binary signal format---qubit, such as the polarization states. Therefore the transmitted information efficiency of the quantum key is intrinsically upper bounded by 1 bit per photon. High dimensional quantum system is a potential candidate for increasing the capacity of single photon. However, due to the difficulty in manipulating and measuring high dimensional quantum systems,  the experimental high dimensional QKD is still at its infancy. Here we propose a sort of practical high-speed high dimensional QKD using partial mutual unbiased basis (PMUB) of photon's orbital angular momentum (OAM). Different from the previous OAM encoding, the high dimensional Hilbert space we used is expanded by the  OAM states with same mode order,  which can be extended to considerably high dimensions and implemented under current state of the art. Because all the OAM states are in the same mode order, the coherence will be well kept after long-distance propagation, and the detection can be achieved by using passive linear optical elements with very high speed. We show that our protocol has high key generation rate and analyze the anti-noise ability under atmospheric turbulence. Furthermore, the security of our protocol based on PMUB is rigorously proved. Our protocol paves a brand new way for the application of photon's OAM in high dimensional QKD field, which can be a breakthrough for high efficiency quantum communications.

\bigskip
\begin{keywords} \textbf{Key words: High-dimensional quantum system, Quantum key distribution,  Photon's orbital angular momentum}
\end{keywords}\bigskip
\end{abstract}

%\address{ Key Laboratory of Quantum Information and Quantum Optoelectronic Devices, Shaanxi Province, Xi'an Jiaotong University, Xi'an 710049, China}
%\address{Center for Quantum Information,  Institute for Interdisciplinary Information Sciences, Tsinghua University, Beijing 100084, China}
%\address{ These authors contributed equally to this work.}

\maketitle

\noindent \textbf{Introduction}

Quantum key distribution (QKD) is one of the best-known applications of quantum information, which promises in principle unconditional secure communications---the Holy Grail of communication security---based on the law of physics only \cite{BB84,E91,3,4,5}. Owing to the quantum non-cloning theorem, QKD system makes it impossible for an eavesdropper to keep a transcript of quantum signals. For  this reason, QKD is an essential element of the future quantum-safe infrastructure.

 A typical QKD protocol involves two parties, conventionally called Alice and Bob, who aim to generate a secret key by exchanging quantum systems over an insecure communication channel \cite{6a,7a,8a,9a,10a,11a,12b,13b,14b,15b}. Security is assessed against the most powerful attack on the channel, where an eavesdropper, conventionally called Eve, perturbs the quantum systems using the most general strategies allowed by physical laws  \cite{16,17,18,19,20}. In general, traditional QKD protocols are performed with qubits, which are two-level quantum systems. In these binary QKD systems, the information efficiency is limited to 1 bit per photon. However, the QKD protocol requires the efficiency of key transmission to be as high as possible. High transmission efficiency allows more data to be encrypted, and the anti-noise performance of quantum channel is improved with the increase of key transmission efficiency.

Over the last decades we have witnessed the advances of  high-dimensional quantum cryptography. The use of high dimensional quantum systems allow for more information to be transmitted between the communicating  parties. And the QKD protocols based on qudit encoding (unit of information in a $d$ dimensional space) exhibit a higher resilience to noise, allowing for lower signal-to-noise ratio of the received signal, which in turn may be translated into higher transmission efficiency and longer transmission distance \cite{23,24,26,27}. As a result, high dimensional QKD (HDQKD) has great potential for developing \cite{26-1,26-2,26-3,26-4,26-5}. Unfortunately, this potential of HDQKD has not been fully fulfilled so far. This is because HDQKD has a critical obstacle in efficient and fast-speed high dimensional quantum states generation and measurement. 

As one of the potential choices for high dimensional quantum system, photon's orbital angular momentum (OAM) has a promising perspective \cite{35,36,37}.  On one hand, the OAM quantum number $l$ can be any arbitrary integer, which corresponds to  infinite dimensional Hilbert spaces \cite{38}. On the other hand, OAM encoded quantum systems are suitable for communication over free-space link due to its resilience against perturbation effects caused by atmospheric turbulence.  Up until now, a number of studies have investigated the benefits employing OAM modes in quantum cryptography \cite{OAM0,OAM1,OAM2,OAM3,OAM4,OAM5}. However, the realization of existing OAM coding HDQKD protocols are still impractical so far mostly due to the difficulty in efficiently measuring single photons in the OAM basis. Because the bases of these protocols are just constructed by different order of OAM states, which will be totally decoherent for long-distance propagation. Furthermore, the repetition rate of the system lies at the range of KHz in all of the existing OAM coding HDQKD, casting serious doubts about their perspective for real world applications.    

Here, we take a major step overcoming the above drawbacks and  propose a sort of practical high-speed HDQKD protocol. In our protocol, the  partial mutual unbiased basis (PMUB) using OAM states with same mode order is constructed for the first time. This  important change leads to stable propagation and easy measurement  for all the OAM states. The generation, manipulation and detection of the same mode order OAM states can be realized by passive linear optical elements. Thus the repetition rate of the system will theoretically reach to GHz. 
By using $\pi/2$ converter, the experimental complexity will not increase greatly with the rise of the dimension. %This advantage is suitable for different demands on aspect of experiments.
\\
\\
\noindent \textbf{Results}

\noindent \textbf{Partial mutual unbiased basis.}
Laguerre-Gaussian (LG) mode beam is a typic one carrying OAM of photon. In this section we introduce expansion formulas for LG  and Hermite-Gaussian (HG) modes, which is the foundation of establishing the PMUBs.  At the single photon level, by using relation between Hermite and Laguerre polynomials, an LG state can be decomposed into a set of HG states of the same order \cite{39}
\begin{equation}
      \ket{l_{nm}}=\sum_{k=0}^{N}i^{k}b(n,m,k)\ket{h_{N-k,k}},
\end{equation}
with real coefficients
\begin{equation}
\begin{aligned}
      b(n,m,k)=&(\dfrac{(N-k)!k!}{2^{N}n!m!})^{1/2}\times       
      \dfrac{1}{k!}\dfrac{d^{k}}{dt^{k}}[(1-t)^{n}(1+t)^{m}]_{t=0},
\end{aligned}
\end{equation}
where $k$ is the wave number, $N=n+m$ is the order of  mode. The factor $i^{k}$ in Eq. (1) corresponds to a $\pi/2$ relative phase difference between successive components. Similarly, an HG state whose principal axes has been rotated $45^\circ$ can be decomposed into exactly the same constituent set
\begin{equation}
\begin{aligned}
      \ket{h_{nm}}=\sum_{k=0}^{N}b(n,m,k)\ket{h_{N-k,k}},
\end{aligned}
\end{equation}
with the same real coefficients $b(n,m,k)$ as above.
\begin{figure}[hbt]
\centering
\includegraphics[width=130mm]{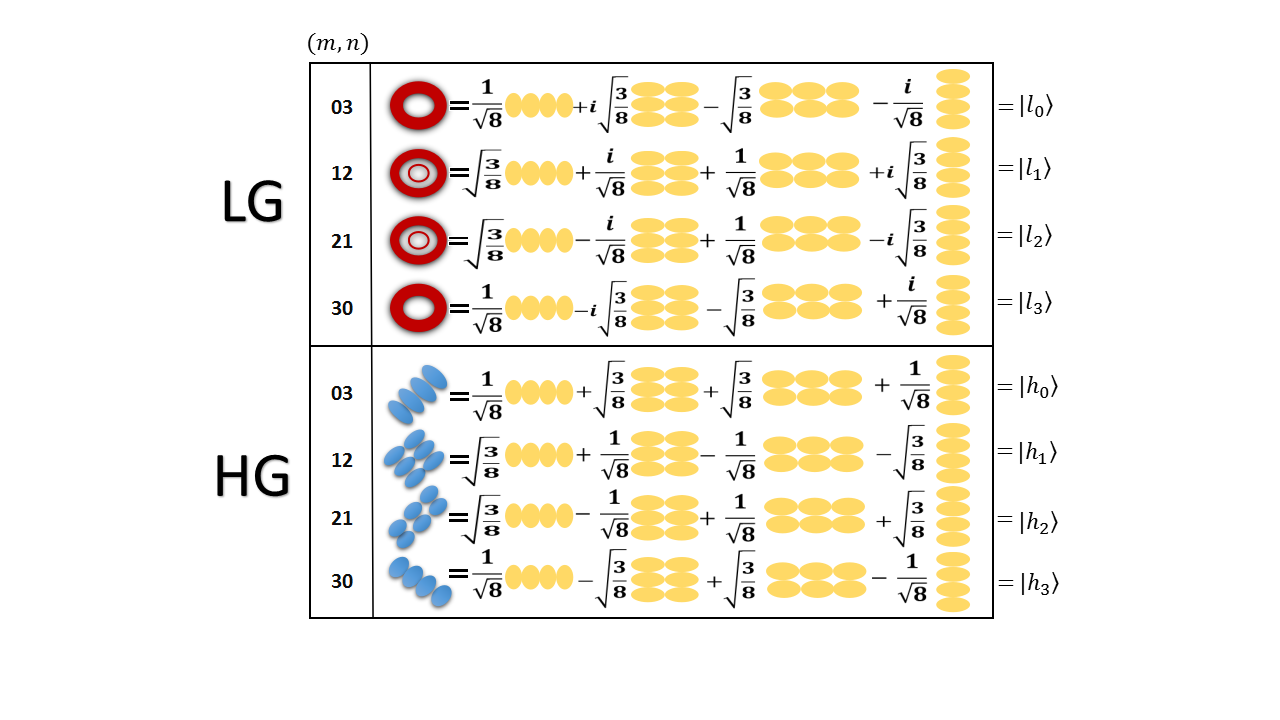}
\caption{Examples of the decomposition of LG (red) and HG (blue) modes of order 3.} 
\label{aad}
\end{figure}

Most of the QKD protocols are based on the concept of mutually unbiased basis (MUB). Because in a set of MUBs $\lbrace B_{0}, B_{1}, B_{2},....., B_{k}, B_{n} \rbrace$, if a state prepared in $B_{k}$ basis is measured in $B_{n}$ basis (with $k\neq n$), all the outputs are equally probable.  However when constructing MUBs in the OAM based HDQKD protocol, the coherence of superposition states between different orders OAM states will be destroyed after long-distance propagation, making it difficult to perform the practical secure key transmission.  

In our protocol,  the problem of decoherence in long-distance propagation is solved by constructing two PMUBs ${l_{nm}}$ (the LG basis) and ${h_{nm}}$ (the HG basis),  which are given by
\begin{subequations} 
\begin{align}
{l_{nm}}=\lbrace  l_{0,N}, l_{1,N-1}, l_{2,N-2},..., l_{N,0} \rbrace,
\\
{h_{nm}}=\lbrace h_{0,N}, h_{1,N-1}, h_{2,N-2},..., h_{N,0} \rbrace,
\end{align}
\end{subequations}
where $N$ can be any positive odd number for the reason of extensibility of the protocol.  The biggest advantage of the chosen two bases is that all the states are in the same mode order. Although ${l_{nm}}$ and ${h_{nm}}$ are not mutually unbiased, according to Eqs. (1)-(4), each individual photon state cannot be fully distinguished in every bases. So our protocol is still secure,  the detailed and rigorous security proof for PMUB is shown in the Method section. 
\\
\\
\noindent \textbf{Four dimensional QKD protocol description.} Based on the above analysis, the HDQKD protocol can be realized based on two PMUBs ${l_{nm}}$ and ${h_{nm}}$ for any positive odd number $N$. For simple and clear description for our HDQKD, here we consider the case of four dimensional QKD protocol with $N=3$ as an example.  Figure 1 shows the states for our protocol (HG and LG modes of order 3). Each time the state Alice chooses for key encoding is one of the eight states in the LG mode (denoted as $\ket{l_i}$, $i=0\sim3$) and the HG mode (denoted as $\ket{h_i}$, $i=0\sim3$).

Denote $\vec{L} = \{\ket{l_0},\ket{l_1},\ket{l_2},\ket{l_3}\}^T, \vec{H} = \{\ket{h_0},\ket{h_1},\ket{h_2},\ket{h_3}\}^T$ as the basis vector, then
%\begin{equation}
%\label{eqn: LUH}
%\vec{L} = U_{LH} \vec{H},
%\end{equation}
$\vec{L} = U_{LH} \vec{H}$ with
$$ U_{LH} = \dfrac{1}{4} 
\begin{pmatrix}
-1+i & \sqrt{3}(1+i) & \sqrt{3}(1-i) & -1-i \\
\sqrt{3}(1+i) & 1-i & 1+i & \sqrt{3}(1-i) \\
\sqrt{3}(1-i) & 1+i & 1-i & \sqrt{3}(1+i) \\
-1-i & \sqrt{3}(1-i) & \sqrt{3}(1+i) & -1+i \\
\end{pmatrix}. $$
Therefore, the four dimensional QKD protocol can be described as:
\begin{enumerate}
\item\label{itm:encode} Alice generates two random bits $a_1,a_2$ (all the random bits we mentioned are generated with uniform possibility distribution) as information to be encoded, and generates one random bit $P_{A}$ for basis choosing. Alice uses $P_{A}$ to determine the encoding basis: $\{\ket{l_i}\}$ or $\{\ket{h_i}\}$, $i=0\sim3$, and uses $a_1,a_2$ to decide which state $\ket{l(h)_i}$ to be sent. Then Alice sends the qudit state to Bob.
\item\label{itm:decode} Bob generates one random bit $P_{B}$ to determine the measurement basis. Upon receiving the state, Bob measures the qudit state on $\{\ket{l_i}\}$ or $\{\ket{h_i}\}$ basis. From the measurement result, Bob receives two bit information $d_1,d_2$.
\item Alice and Bob do steps \ref{itm:encode}$\sim$\ref{itm:decode} many rounds and keep $a_1,a_2$, $d_1,d_2$, $P_{A}$, $P_{B}$ as raw data for later use. We denote $a=2a_1+a_2$, $d=2d_1+d_2$, thus $0\leq a,d \leq 3$.
\item Sifting process: Alice and Bob announce and compare all the $P_{A}$, $P_{B}$ data. They compare $P_{A}$ and $P_{B}$, and throw all the corresponding raw data where $P_{A} \neq P_{B}$ and keep the $a,d$ with $P_{A}= P_{B}$ as the raw key.
%\item Parameter estimation: Alice and Bob randomly sample part of raw key and announce them. By comparing the raw key, they obtain the error syndrome: for the case $b=c=0$ (i.e., the basis $\ket{l_i}$), suppose we have $S^{(0)}$ samples, denote $S^{(0)}_{ad}$ as the sample number of the case where Alice's raw key is $a$ and Bob's raw key is $d$. Define $p_{ad}^{(0)} = \dfrac{S^{(0)}_{ad}}{S^{(0)}}$. Similarly, for $b=c=1$ case, we can denote $S^{(1)}, S^{(1)}_{ad}$ and define $p_{ad}^{(1)} = \dfrac{S^{(1)}_{ad}}{S^{(1)}}$. Alice and Bob use $\{p^{(b)}_{ad}\}\; (b=0,1; 0\leq a,d \leq 3)$ to determine the key rate $r$ and the number of key needed $k_{ec}, k_{pa}$ to perform error correction and phase estimation, respectively.
\item Alice and Bob perform standard post-processing method to generate secure and identical key.
\end{enumerate}

\noindent \textbf{Experiential approach.} 
Figure 2 shows the principle experiment sketch of our four-dimensional QKD protocol. The LG state generators are used to prepare the original LG$^{p}_{l}$ states  with  parameters $l$ and $p$ (the azimuthal index $l$ is $n-m$ and the radial index $p$ equals to $\min[n,m]$), and the main component is a spatial light modulator.  An intensity modulator is used to generate decoy states. The decoy-state method \cite{D1,D2,D3,D4,D5} can be used to protect the transmission process from  the photon number-splitting attack \cite{P1,P2,P3,E1,E2,E3}.  %According to Fig.1 and the relationships between parameters ($n$, $m$) and ($p$, $l$), Alice prepares the states LG$^{0}_{4}$, LG$^{0}_{0}$, LG$^{1}_{4}$ and LG$^{1}_{0}$.   
All the generators are  controlled by acoustic-optical modulators. In the state preparation and measurement part, the interferometric method in reference \cite{M1} is used to combine and sort the LG states. The  Mach-Zehnder interferometers with two Dove prisms placed in each arm, and the relative angle between the Dove prisms is $\alpha/2$. In the first and forth stages, $\alpha=\pi/4$ corresponds to a relative phase difference $\Delta \Psi=l\pi/4$ between the two arms of the interferometers. Therefore,  states with $l=8\omega$ and $l=8\omega+4$ come out in different ports, where $\omega$ is an integer. Similarly, in the second and third stages,  $\alpha=\pi/2$ corresponds to $\Delta \Psi=l\pi/2$, states with $l=4\omega$ and $l=4\omega+2$ come out in different ports.
Spiral phase plates in each stages are used to displace OAM $l$ of the photons. Therefore, when Alice prepares the states LG$^{0}_{4}$, LG$^{0}_{0}$, LG$^{1}_{4}$ and LG$^{1}_{0}$ at the beginning, the states entered in the transmission channel will be  LG$^{0}_{3}$, LG$^{0}_{-3}$, LG$^{1}_{1}$ and LG$^{1}_{-1}$ as the protocol required. With the help of $\pi/2$ converter, LG mode states and HG mode states can be transformed easily.
 
%This kind of experiential approach avoid using random gratings (the traditional method of OAM based QKD for generating LG states). The repetition rate is mainly  based on  acoustic-optical modulators, which can be maintained at the range of GHz. However, there are also two practical issues that should be addressed. Firstly, a $\pi/2$ converter requires incident light having certain wavelength. Secondly, a high order LG sorter requires a cascade of interferometers. In practical experiments and applications, researchers and users  can determine the order of the protocol according to the their concrete requirements, taking both the transmission efficiency and the complexity of experiment into account.

\begin{figure}[hbt]
\centering
\includegraphics[width=150mm]{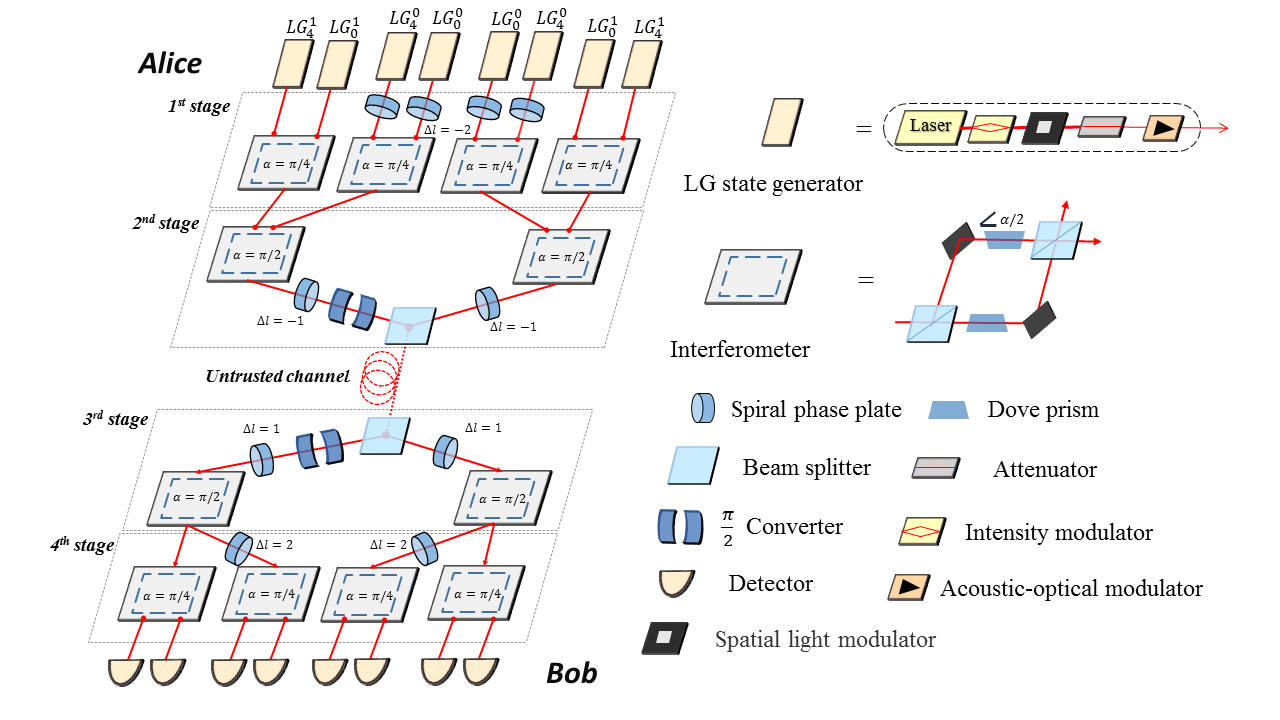}
\caption{ Experiential approach for four-dimensional QKD.  The first and forth stages introduce a phase shift of $\alpha=\pi/4$. The second and third stages introduce a phase shift of $\alpha=\pi/2$.  The displacement of OAM introduced by spiral phase plates is $\Delta l=-2$ in the first stage, $\Delta l=-1$ in the second stage, $\Delta l=1$ in the third stage, and $\Delta l=2$ in the forth stage. Two $\pi/2$ converters are used for the transformation between LG mode and HG mode. } 
\label{tu}
\end{figure}

\noindent \textbf{Security  key rate based on numerical method.}
Typically, Alice's and Bob's shared density operater $\rho_{AB}$ is unknown to them. They gather data through local measurements and use the data to constrain the form of $\rho_{AB}$. Following the methods in reference \cite{46}, the measurements  can be described by a set of bounded  Hermitian operators $\vec{\Gamma}=\{\Gamma_i\}$. From their data, Alice and Bob determine the average value of each of the measurements,

\begin{equation}
\label{eqn: le}
\begin{aligned}
 \quad \gamma_{i}=\langle \Gamma_i \rangle=\text{Tr}(\rho_{AB} \Gamma_i),
\end{aligned}
\end{equation}
which gives a set of experimental constraints,
\begin{equation}
\label{eqn: ll}
\begin{aligned}
\lbrace \text{Tr}(\rho_{AB} \Gamma_i)=\gamma_{i} \rbrace
\end{aligned},
\end{equation}
and an additional constraint $\langle I\rangle =1$ is assumed to  this set  to enforce normalization. And according to \cite{46},  key rate is given  by the following maximization problem:
\begin{equation}
\label{eqn: l2}
\begin{aligned}
K \geq \frac{\kappa}{\ln 2}-H(M_L(A)|M_L(B),
\end{aligned}
\end{equation}
where where $H(X|Y):= H(\rho_{XY}) - H(\rho_Y)$ is the conditional Von Neumann entropy, with $H(\sigma):=-Tr(\sigma\log_2{\sigma})$,  $M_L(A)$  is the measurement on basis $\{\ket{l_i^{(\prime)}}\}$ (the basis of the equivalent entanglement-based protocol, see details in the method part), 
\begin{equation}
\label{eqn: l3}
\begin{aligned}
\kappa=\max (-\Vert \sum_{j}M_L(A)^{j} T(\vec{\lambda}) M_L(A)^{j}- \vec{\lambda}\cdot \vec{\gamma}\Vert),
\end{aligned}
\end{equation}
and
\begin{equation}
\label{eqn: l4}
\begin{aligned}
 T(\vec{\lambda})=\exp (-I-\vec{\lambda}\cdot \vec{\Gamma}).
\end{aligned}
\end{equation}
In Eqs. (8) and (9), the optimization is over all vectors $\vec{\lambda}=\lbrace \lambda_{i}\rbrace$, where $\lambda_{i}$ are arbitrary real numbers, $\lambda$ and 
$\vec{\Gamma}$ have equal cardinality. In the PMUBs case, the corresponding  constraints are given by
\begin{equation}
\begin{aligned}
&\langle I\rangle=1, \\
&\langle M_L\otimes M_L\rangle=1-2Q, \\
&\langle M_H\otimes M_H\rangle=1-2Q, \\
&\langle M_L\otimes M_H\rangle=(\sin \theta)(1-2Q), \\
&\langle M_H\otimes M_L\rangle=(\sin \theta)(1-2Q), \\
\end{aligned}
\end{equation}
where $\theta=\max \lbrace\arccos \langle l_{i}\vert h_{i}\rangle\rbrace$, $M_L$ and $M_H$ are the measurements on basis $\{\ket{l_i}\rbrace$ and $\{\ket{h_i}\rbrace$.
The biggest advantage of this numerical method is that the number of parameters one is optimizing over just equal to the number of constraints (in this case is 5), which is independent of dimension. Figure 3 plots the key rate of our four dimensional QKD  and BB84 protocols as a function of the error rate $Q$, which show the fact that the variations in $\theta$ have essentially no effect on the key rate in the case of our protocol and shows it has  better error tolerance. 
\begin{figure}[hbt]
\centering
\includegraphics[width=130mm]{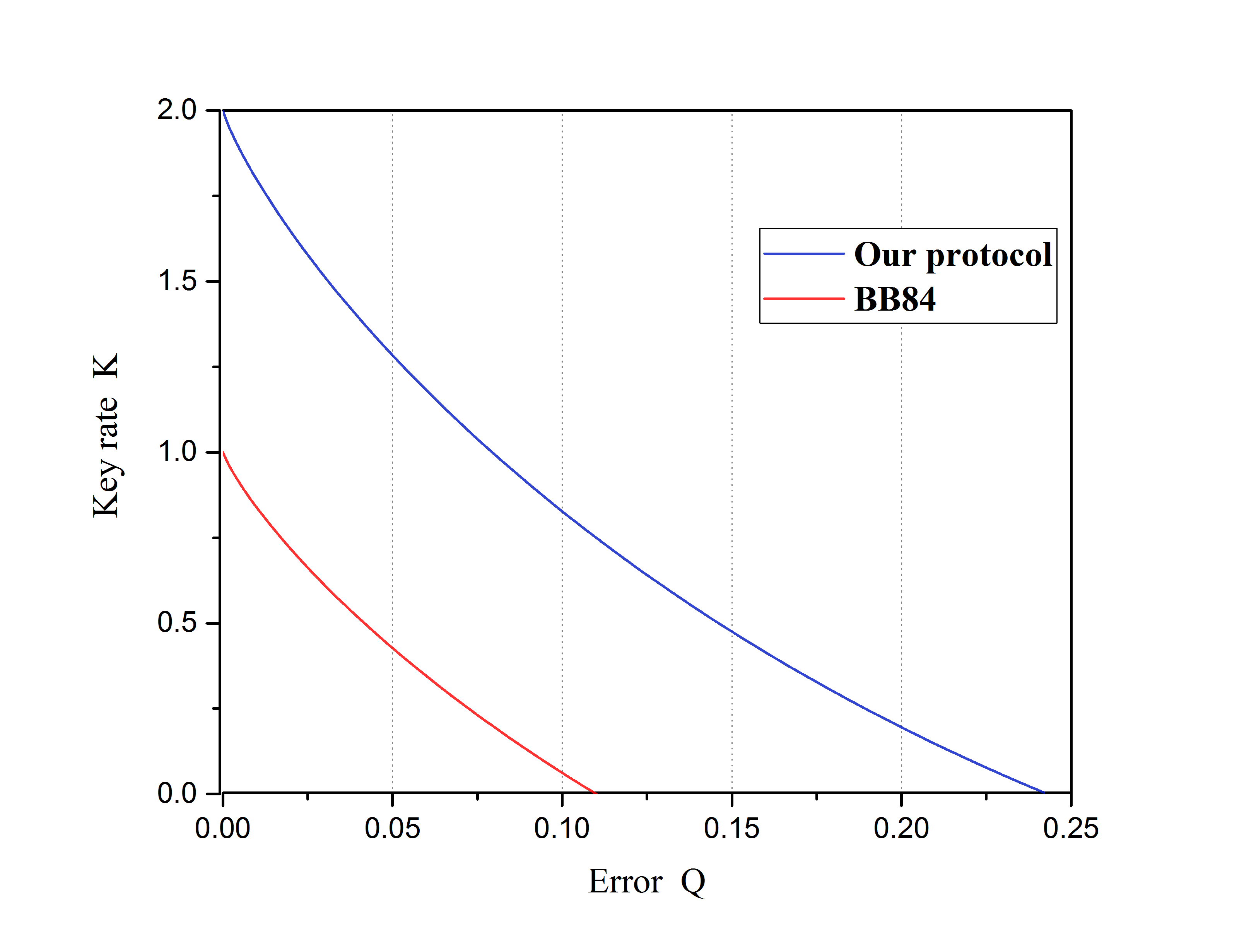}
\caption{Key rate of four dimensional QKD (blue line)  and BB84 (red line) based on dual optimization method.} 
\label{fig:G}
\end{figure}

\noindent \textbf{Practical key rate based on turbulence model.}
\begin{figure}[hbt]
\centering
\includegraphics[width=130mm]{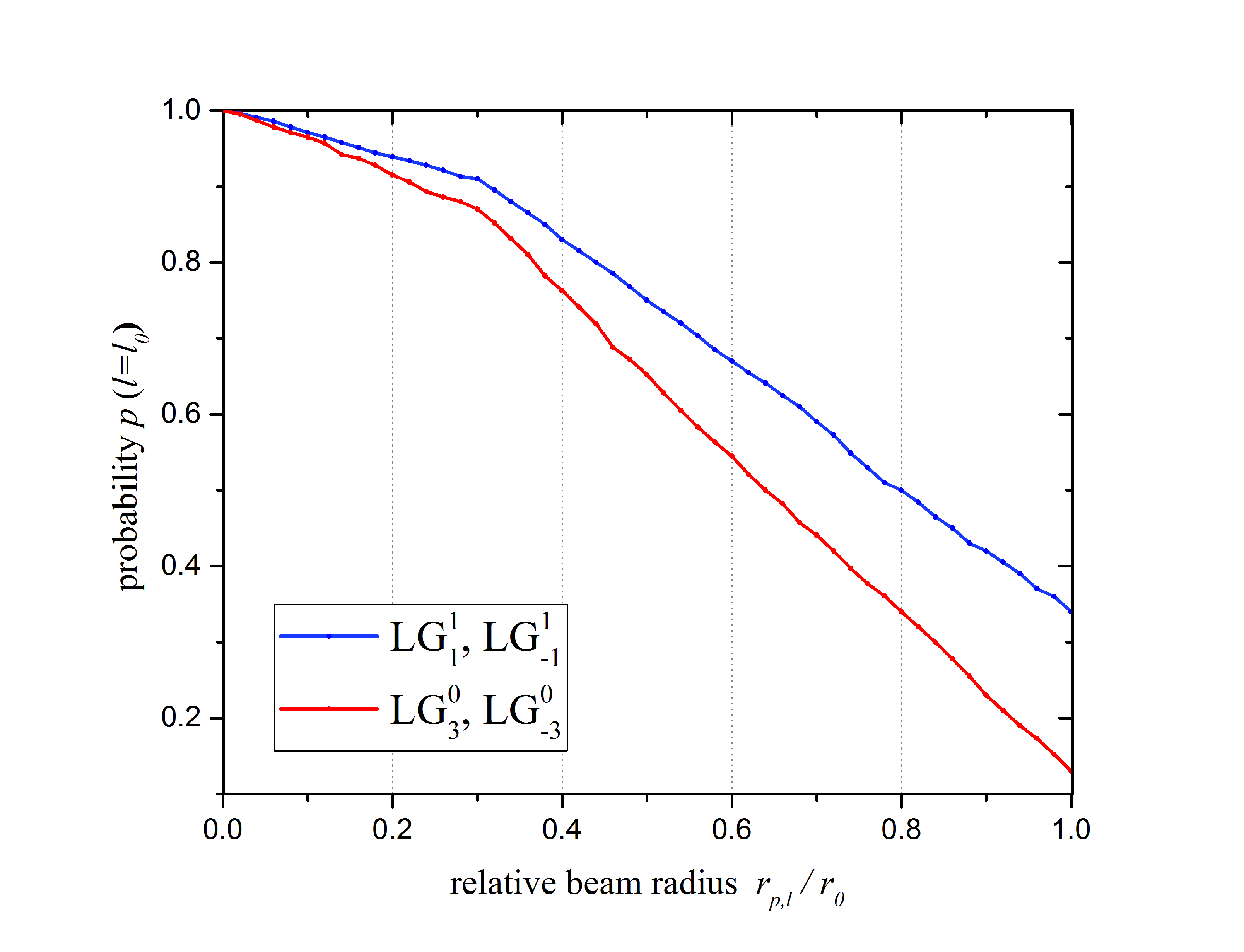}
\caption{ Probabilities of obtaining the original eigenvalue
for the four LG  modes plotted against the ratio of the relative mean-squared beam radius to the Fried parameter.} 
\label{dada}
\end{figure}
For the LG channel, after approximating the cumulative effect of the turbulence over the propagation path as a pure phase perturbation on the beam at the output plane $z$, the conditional probability of obtaining a measurement of the OAM of a photon $l_{z}=l\hbar$ is given by
\begin{equation}
      p(l)=\int_{0}^{\infty} \vert R(r,z) \vert^{2}r\Theta(r,l-l_{0})dr, 
\end{equation}
where $\Theta(r,\Delta l)$ is the circular harmonic transform of the rotational coherence function, which is given by
\begin{equation}
     \Theta(r,\Delta l)=\dfrac{1}{2\pi}\int_{0}^{2\pi}C_{\phi}(r, \Delta \theta) \exp [-i\Delta l\Delta \theta]d\Delta \theta,  
\end{equation}
where $C_{\phi}(r, \Delta\theta)$ is the rotational coherence function of the phase perturbations at radius $r$. For Kolmogorov turbulence phase statistics, the rotational coherence function at radius $r$ is
\begin{equation}
     C_{\phi}(r, \Delta\theta)=\exp [-6.88\times 2^{2/3}(\dfrac{r}{r_{0}})^{5/3} \vert \sin(\dfrac{\Delta \theta}{2}) \vert^{5/3}],
\end{equation}
where $r_{0}$ is the Fried parameter \cite{L1}. The OAM probabilities for various LG states propagating through Kolmogorov turbulence can be evaluated using Eqs. (11)-(13). For different order state modes, the effect of the phase perturbations depends on the radial power distribution in the beam which for an LG$_{l}^{p}$ is
\begin{equation}
\langle r^{2} \rangle=\int^{\infty} _{r=0}R_{l,p}(r)r^{2}dr=(2p+l+1)b^{2}, 
\end{equation}
giving a characteristic relative mean-squared beam radius $r_{p,l}=b\sqrt{2p+l+1}$. 
For the HG channel, during the measurement process, the HG states are converted to corresponding LG states by using the $\pi/2$ converter. What's more, in the above analysis, the cumulative effect of the turbulence over the propagation path is considered as a pure phase perturbation $\exp(i\phi)$ on the beam at the output plane $z$. Therefore, the bit error rate can be  calculated in the same way above.

Figure 4 plots the probabilities for obtaining the original OAM eigenvalue scaled against the  relative mean-squared beam radius for the LG$^{0}_{3}$, LG$^{0}_{-3}$, LG$^{1}_{1}$ and LG$^{1}_{-1}$ states. When $b=0.01$ m (corresponds to moderate ground-level turbulence strength $10^{-14}$m$^{-2/3}$ and wavelength $\lambda=1 \mu$ m),  this  probability  is evaluated as $p (l=l_{0}) = 0.88\pm 0.051$.  Regarding the average bit error rate $Q$ as a statistical average composed of the bit error rate in LG and HG channel, the bit error rate is obtained as $Q\geq0.12\pm 0.051$. The statistical errors of $p (l=l_{0})$ and $Q$  come from different chooses of $(n, m)$.

  Figure 5 plots the simulation results that the practical security key rate $K_{P}$ as a function of time of our four dimenional QKD (blue line) and polarization coding reference-frame-independent QKD (red line) which has been experimental demonstrated to be robust for slowing varying reference frames \cite{R1,R2,R4,R5,R6,R7}. For the reference-frame-independent QKD  protocol, the noise parameter  and the sampling interval are seted as $0.5$s and $\pi$  according to the result of \cite{R8}. It is easy to notice that the statistical errors of the two protocol are close, which highlights that our protocol has  anti-noise ability. 
  
\begin{figure}[hbt]
\centering
\includegraphics[width=130mm]{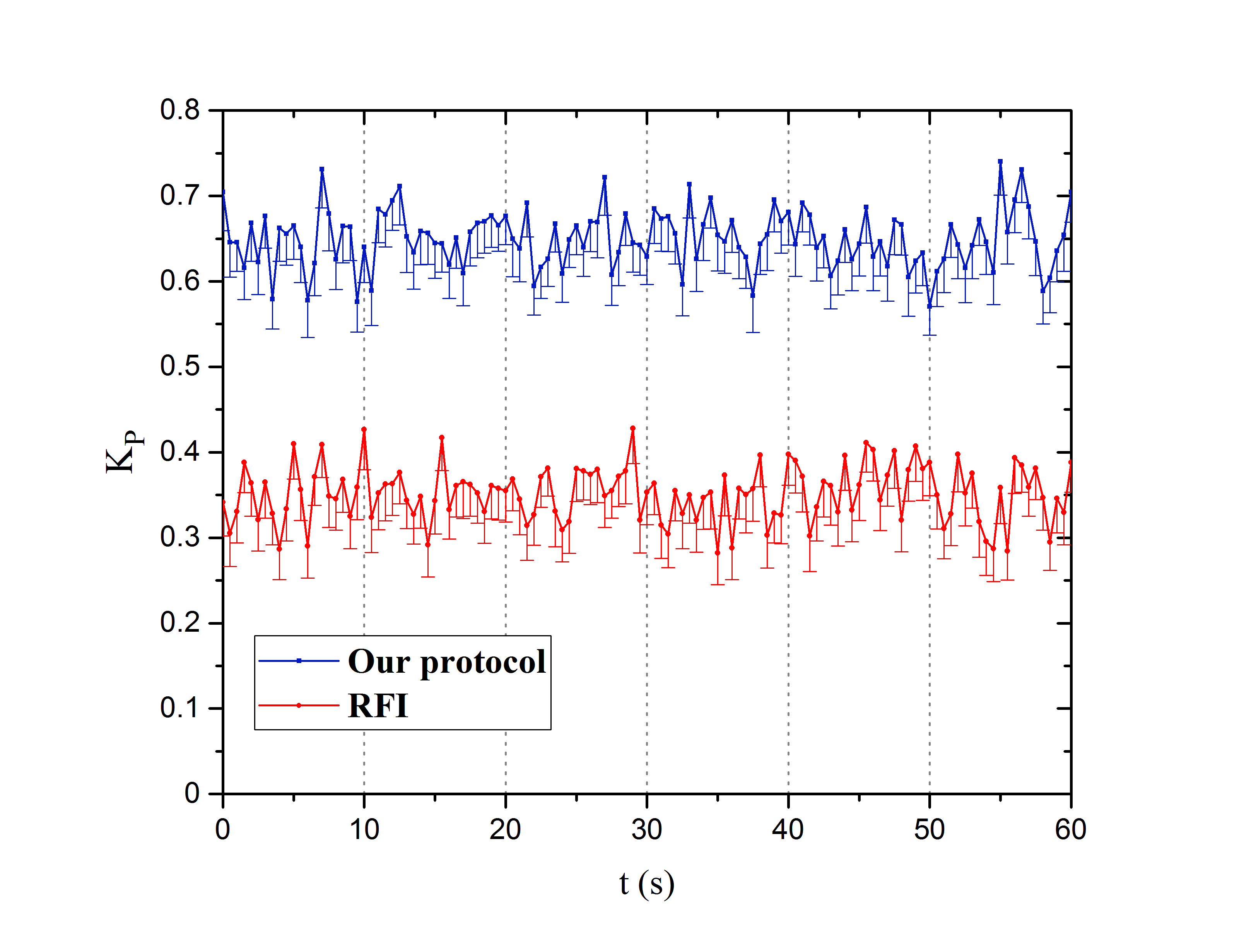}
\caption{Simulation results that the practical security key rate $K_{P}$ as a function of time of our protocol (blue line) and polarization coding reference-frame-independent QKD protocol (red line). } 
\label{KS}
\end{figure}

\noindent \textbf{Discussion}

In summary, we have proposed a sort of high dimensional QKD protocol for practical application.  In theory, we use  PMUBs to overcome states transmission and measurement issues so as to improve the security key rate.  For experimental realization, a detailed approach just using passive devices is designed, in which the  state generation  mainly depends on acoustic-optical modulators, resulting in a repetition rate at the range of GHz which is comparable with phase encoding  protocol.  Moreover, our protocol has  extensibility, which means one can use higher dimensional states for key encoding. In view of its good security and reasonable implementation, we believe that our protocol will be a big step forward for high information efficiency quantum communication.

\noindent \textbf{Method}

\noindent \textbf{Security analysis based on uncertainty relationship.}
Suppose there are two Hermitian operators $X$ and $Z$ on the $L$-dimension Hilbert space $\mathcal{H}$, the corresponding orthonormal basis groups are $\{\ket{x_i}\}, \{\ket{z_i}\}\,(i=1\sim L)$. For an arbitrary state $\rho\in\mathcal{H}$, the project-value measurement results on two orthonormal basis are denoted as: $\{p^{(x)}_i\}, \{p^{(z)}_i\}\, (i=1\sim L)$. Denote the Shannon entropy of measurement results as: $H_X(\rho), H_Z(\rho)$. From the work \cite{40, 41, 42} and \cite{43}, the  entropy uncertainty relationship is given by
\begin{equation}
\label{eqn: uncertain}
H_X(\rho) + H_Z(\rho) \geq \log(\dfrac{1}{c}) := q_{MU} \quad \forall \rho\in \mathcal{H},
\end{equation}
where $c$ is defined as the maximum overlap of two basis
\begin{equation}
\label{eqn: c}
c = \max_{i,j} c_{i,j}, \quad c_{i,j}:=|\inner{x_i}{z_j}|^2,
\end{equation}
and $q_{MU} = \log(\dfrac{1}{c})$. For the two orthonormal basis $ \{\ket{l_i}\} (i=0\sim3)$ or $\{\ket{h_i}\} (i=0\sim3)$, the maximum overlap is given by $ c = \max_{i,j} c_{i,j} = \dfrac{3}{8}$, therefore $q_{MU} = \log(\dfrac{1}{c}) = 3-\log{3}. $
Hence the entropic uncertainty relationship for $\{\ket{l_i}\}, \{\ket{h_i}\}$ is given by 
\begin{equation}
\label{eqn: lh_uncertain}
H_{LG}(\rho) + H_{HG}(\rho) \geq 3-\log{3} \quad \forall \rho\in \mathcal{H}.
\end{equation}

For the security analysis,  a tripartite uncertainty relations is often needed. In a tripartite scenario (as shown in Fig.6), state $\rho_{ABE}$ is divided into three parts $A, B, E$ and sent to Alice, Bob and Eve, respectively. Suppose the subsystem held by Alice is $\rho_A$, and there are two complementary measurement bases ($X$ and $Z$) for $\rho_A$. Alice perform measurement on either $X$ or $Z$ basis. If Alice measures $X$, then Bob's goal is to minimize his uncertainty on $X$ measurement result $H(X_A|B)$; if she measures $Z$, then Eve's goal is to minimize his uncertainty on $Z$ measurement result $H(Z_A|E)$. In this case, Bob and Eve hold quantum systems $\rho_B,\rho_E$, making them able to choose proper measurement basis to optimize their knowledge on Alice's measurement, thus Von Neumann entropy here is a lower bound over all the possible measurements on $\rho_B$ or $\rho_E$.
Renes et,al. \cite{44} shows that there exists uncertainty relationship on $H(X_A|B), H(Z_A|E)$
\begin{equation}
\label{eqn: 3P_uncertain}
H(X_A|B) + H(Z_A|E) \geq q_{MU}.
\end{equation}

\begin{figure}[hbt]
\centering
\includegraphics[width=0.8\textwidth]{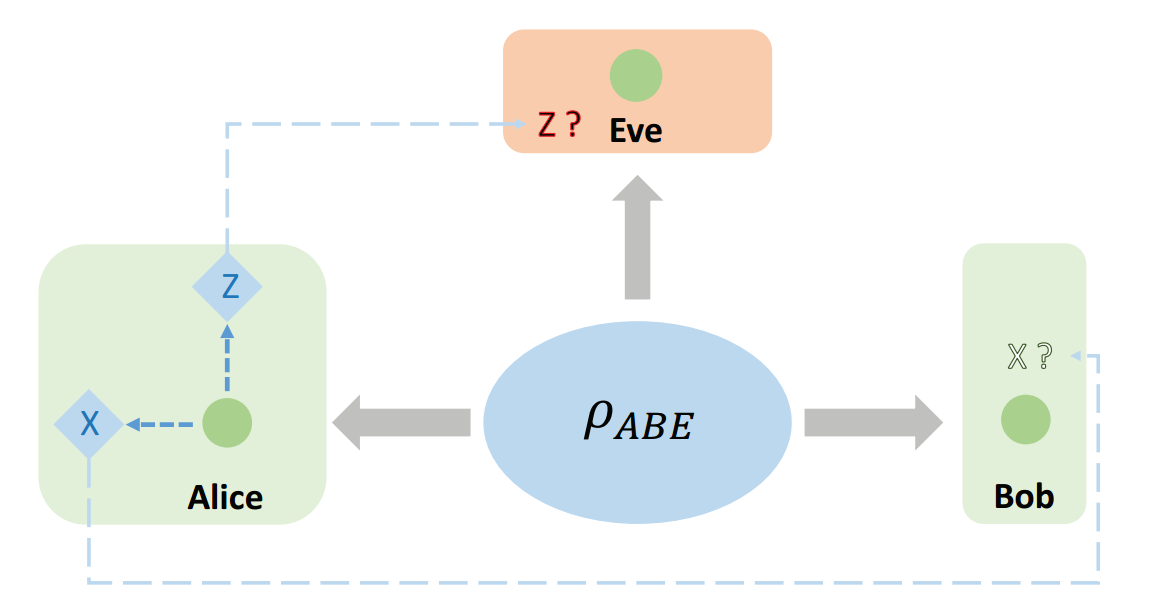}
\caption{Scenario of tripartite uncertainty relationship.} 
\label{fig:3P_uncertain}
\end{figure}

We take Devetak-Winter's approach \cite{45} for security analysis, which is based on the entanglement distillation of a entanglement-based QKD protocol. For our BB84-like protocol,   a equivalent entanglement-based protocol can be easily defined. Two protocols are equivalent with respect to Eve if and only if:
\begin{enumerate}
	\item The transferred quantum state from Alice and all the classical signal revealed are the same;
	\item Bob's measurement result statistics are the same.
\end{enumerate}
Suppose Alice prepare state
\begin{equation}
\label{eqn: AliceState}
\rho_0 := \dfrac{1}{2} \sum_{i=0}^3 {\ket{l_i^\prime}}_A \ket{l_i} = \dfrac{1}{2} \sum_{i=0}^3 {\ket{h_i^\prime}}_A \ket{h_i}
\end{equation}
which is defined on Hilbert space $\mathcal{H}_A \otimes \mathcal{H}$. The qudit on space $\mathcal{H}_A$ is the ancillary bit kept by Alice which is used to determine the encoded information. Basis $\{\ket{l_i^\prime}_A\}$ and $\{\ket{h_i^\prime}_A\}$ are two orthonormal basis on $\mathcal{H}_A$ which promise eq.~\eqref{eqn: AliceState} holds. 

Denote $\vec{L^\prime} = \{\ket{l_0^\prime},\ket{l_1^\prime},\ket{l_2^\prime},\ket{l_3^\prime}\}^T, \vec{H^\prime} = \{\ket{h_0^\prime},\ket{h_1^\prime},\ket{h_2^\prime},\ket{h_3^\prime}\}^T$. From Eq.~\eqref{eqn: AliceState} we can obtain

\begin{equation}
\vec{H^\prime} = U_{LH} \vec{L^\prime} \Rightarrow \vec{L^\prime} = U_{LH}^{-1} \vec{H^\prime}.
\end{equation}

Alice then randomly choose to measure her ancillary qudit on $\mathcal{H}_A$ in basis $\{\ket{l_i^\prime}_A\}$ or $\{\ket{h_i^\prime}_A\}$ based on the random number $P_{A}$ she generates. She keeps her measurement result as raw data $a$ and sends the qudit in space $\mathcal{H}$ to Bob. Bob performs the same measurement as described in the real protocol. It's easy to show that this entanglement-based protocol is equivalent to the real one with respect to Eve. So asymptotic key rate $K_{a}$ for the entanglement-based protocol can be given by the Devatak-Winter formula\cite{45}
\begin{equation}
\label{eqn: Dev-Win Formula}
K_{a} = H(M_L(A)|E) - H(M_L(A)|M_L(B)),
\end{equation}
 and
\begin{align}
\rho_{M_L(A)M_L(B)}  & = \sum_{j,k} Tr[(M_L^j \otimes M_L^k)\rho_{AB}] \ket{l^\prime_j}\bra{l^\prime_j} \otimes \ket{l_k}\bra{l_k},  \\
\rho_{M_L(A)E} & = \sum_{j} \ket{l^\prime_j}\bra{l^\prime_j}\otimes Tr_A [(M_L^j \otimes I)\rho_{AE}].
\end{align}
Term $H(M_L(A)|M_L(B))$ in Eq.~\eqref{eqn: Dev-Win Formula} reflects the cost for classical error correction, which is equal to the classical conditional Shannon entropy of the measurement results $M_L(A), M_L(B)$. Recall the tri-partite uncertainty relationship (eq.~\eqref{eqn: 3P_uncertain}), we have
\begin{equation}
\label{eqn: Mlh_3P_uncertain}
H(M_L(A)|E) + H(M_H(A)|B) \geq q_{MU} = 3 - \log{3},	
\end{equation}
where $q_{MU}$ defined in Eq.~\ref{eqn: uncertain} can be calculated by the basis transform matrix $U_{LH}$. According to Eq.~\eqref{eqn: Dev-Win Formula} and Eq.~\eqref{eqn: Mlh_3P_uncertain}, we can obtain
\begin{equation}
\label{eqn: keyrate}
\begin{aligned}
K_{a} &  \geq (3 - \log{3}) - H(M_H(A)|M_H(B)) - H(M_L(A)|M_L(B)).
\end{aligned}
\end{equation}

\noindent
\\
\noindent \textbf{Acknowledgements}

\noindent We wish to thank professor Xiongfeng Ma from Tsinghua University for helpful discussion. The authors acknowledge support from the Fundamental Research Funds for the Central Universities, Joint Funds of the Ministry of Education of China (Grant No. 6141A02011604), Natural Science Foundation of Shaanxi Province (Grant No. 2017JM6011), and National Natural Science Foundation of China ( Grant Nos. 91736104, 11374008, and 11534008).
\\
\\
\noindent \textbf{Author contributions}

\noindent Pei Zhang conceived the idea. Pei Zhang and Fumin Wang devised and designed the protocol. Pei Zeng and Fumin Wang analyzed the  security of the protocol, Xiaoli Wang, Hong Gao and Fuli Li analyzed the theoretical and experimental approaches. All authors contributed to the writing of the manuscript.
\\
\\
\noindent  \textbf{Additional information}

\noindent  The authors declare no competing financial interests. Reprints and permissions information is available online at http://npg.nature.com/reprintsandpermissions. Correspondence and requests for materials should be addressed to Pei Zhang. %\blk

\end{document}